\documentclass{aa}
\usepackage{graphicx}
\usepackage{graphics}
\usepackage{lscape}

\def\ltsima{$\; \buildrel < \over \sim \;$}
\def\simlt{\lower.5ex\hbox{\ltsima}}
\def\gtsima{$\; \buildrel > \over \sim \;$}
\def\simgt{\lower.5ex\hbox{\gtsima}}

\def\cgs{ ${\rm erg~cm}^{-2}~{\rm s}^{-1}$ } 
\def\ROSAT{{\it ROSAT} }
\def\SAX{{\it BeppoSAX} }
\def\ASCA{{\it ASCA} }
\def\pn{\par\noindent}

\begin{document}
   \title{The BeppoSAX High Energy Large Area Survey.}

   \subtitle{IV. On the soft X--ray properties of the hard X--ray-selected HELLAS sources}

   \titlerunning{Soft X--ray properties of the HELLAS sources}
   \authorrunning{C. Vignali et al.}

   \author{C. Vignali\inst{1,2}
          \and
	  A. Comastri\inst{2}
	  \and
	  F. Fiore\inst{3,4}
	  \and
	  F. La Franca\inst{5}
          }

   \offprints{C. Vignali, \email{vignali@kennet.bo.astro.it}}

   \institute{Dipartimento di Astronomia, Universit\`a di Bologna, 
              Via Ranzani 1, I--40127 Bologna, Italy \\
          \and
                Osservatorio Astronomico di Bologna, 
                Via Ranzani 1, I--40127 Bologna, Italy \\
          \and
                Osservatorio Astronomico di Roma, 
                Via Frascati 33, I--00044 Monteporzio, Italy \\
          \and
	        BeppoSAX Science Data Center, 
                Via Corcolle 19, I--00131 Roma, Italy \\
          \and
                Dipartimento di Fisica, Universit\`a degli Studi ``Roma Tre'', 
                Via della Vasca Navale 84, I--00146 Roma, Italy \\
             }

   \date{}

\abstract{ 
We present a comprehensive study of the soft X--ray properties of the \SAX  High-Energy 
Large Area Survey (HELLAS) sources. 
A large fraction (about 2/3) of the hard X--ray selected sources is detected by {\it ROSAT}. 
The soft X--ray colors for many of these objects, along with the 0.5--2 keV flux upper limits for 
those undetected in the \ROSAT band, do imply the presence of absorption. 
The comparison with the \ROSAT  Deep Survey sources indicates that a 
larger fraction of absorbed objects among the HELLAS sources is present, 
in agreement with their hard X--ray selection and the predictions of the X--ray 
background synthesis models. 
Another striking result is the presence of a soft (additional) X--ray component in a significant 
fraction of absorbed objects. 
}
\maketitle

\section{Introduction}

While a relevant fraction (about 70--80 \%) of the soft X--ray background (hereafter XRB) 
has been resolved into discrete sources 
by the \ROSAT satellite (Hasinger et al. 1993, 1998), most of which being 
broad-line active galactic nuclei (AGNs) 
(Shanks et al. 1991; Boyle et al. 1994; 
Schmidt et al. 1998; Lehmann et al. 2000), 
in the hard band, where the bulk of the energy density resides, 
the nature of the sources of the XRB is far less obvious. 
Before the advent of the imaging instruments onboard 
{\it ASCA} and {\it BeppoSAX}, surveys in the hard energy range have been 
performed with limited-spatial resolution instruments, 
thus allowing the identification of the X--ray brightest sources, which account for 
a small fraction (about 3--5 \%, 
Piccinotti et al. 1982) of the hard XRB. 
The AGNs observed at high energies by {\it HEAO1}, {\it EXOSAT} and 
{\it Ginga} have spectra much steeper (with a photon index $\Gamma$$\simeq$1.7--1.9) 
than the slope of the XRB in the same energy range 
($\Gamma$$\simeq$1.4--1.5, Gendreau et al. 1995; Vecchi et al. 1999). 
This fact, called ``spectral paradox'', has been 
theoretically solved by assuming that the XRB is due to a mixture 
of absorbed and unabsorbed objects (Setti \& Woltjer 1989). 
Following this indication, 
the contribution of different classes of sources to the XRB 
has been evaluated through population-synthesis models (e.g. Madau et al. 1994; Comastri et al. 1995).

Recently, the {\it ASCA} 2--10 keV surveys (Della Ceca et al. 1999; 
Ueda et al. 1999) and the {\it BeppoSAX} {\bf H}igh {\bf E}nergy {\bf LL}arge 
{\bf A}rea {\bf S}urvey (HELLAS) (Fiore et al. 1999, 2000a), 
carried out in the 5--10 keV energy range down to a 
flux limit of about 5$\times$10$^{-14}$ \cgs, have proven to be very efficient 
in revealing the nature of the sources which resolve about 20--30 \% 
of the hard XRB, obscured AGNs being the main contributors. 
{\it Chandra} observations have confirmed and 
extended down to lower fluxes these results (Mushotzky et al. 2000; 
Fiore et al. 2000b; Brandt et al. 2000; Giacconi et al. 2001; Barger et al. 2001). 

In order to better understand the nature and the properties 
of the sources responsible for the hard XRB and to verify whether additional 
soft X--ray components may be present, 
we have systematically searched in the {\it ROSAT} archive 
for complementary X--ray data to the present hard X--ray selected sample. 
Generally, soft X--ray components in addition to the standard AGNs spectral models 
are either excluded from the XRB synthesis models 
(e.g. Comastri et al. 1995) or, if included, 
they do account for a small, energetically not relevant fraction of the XRB (Gilli et al. 1999). 
Indeed, there is increasing evidence that 
the broad-band spectral properties of the sources responsible for the hard XRB are 
characterized by more complex spectra than is expected from a simple absorbed power-law model 
(Della Ceca et al. 1999; Giommi et al. 2000), as confirmed by the presence of 
multi-components spectra in obscured nearby AGNs (Awaki et al. 2000). 
The possibility of the existence of such additional components 
must be taken into account when comparing surveys performed in different energy 
ranges and, most important, at limiting fluxes differing by one (or more) order of 
magnitude.

\section{The sample}

\subsection{The HELLAS hard X--ray-selected sample}

142 high Galactic latitude ($|$b$|$$>$20$^{\circ}$) MECS 
fields covering about 80 square degrees of sky have been analyzed. 
A detailed description of \SAX  instrumental capabilities, source detection algorithms 
and photon statistics is described by Fiore et al. (2001, paper~II). 
For the purposes of the present paper, we have considered the full list of 147 \SAX sources 
detected down to a flux limit of about 5 $\times$ 10$^{-14}$ \cgs.

\subsection{The HELLAS sources in \ROSAT archival data}

The \SAX catalogue 
was cross-correlated with archival {\it ROSAT} data (PSPC, HRI and RASS). 
108 out of the 147 sources are in {\it ROSAT} fields; 
for only 1 source RASS data have been used, due to the lack 
of pointed {\it ROSAT} observations. 
4 sources lying beneath the PSPC detector window 
support structure will not be considered in the following discussions, 
therefore the useful number of HELLAS sources in \ROSAT fields is {\sf 104}. 
In those cases where multiple observations of the same HELLAS 
source are present, we chose the one with the longest exposure time and/or the lowest 
off-axis angle (and checked the results by analyzing at least another 
image of the same field).

\section{ROSAT fields}

\subsection{ROSAT source detection}

The {\it ROSAT} data have been analyzed with the MIDAS/EXSAS package 
(Zimmermann et al. 1998). 
The sources have been detected by running the local 
detection algorithm {\sc LDETECT}, the bicubic spline fit 
to the background map and the map detection algorithm {\sc MDETECT}. 
The detection threshold of these algorithms was set at a likelihood  of 
L=$-$ln(P$_{\rm e}$)=10, corresponding to a probability of the order of 4.5$\times$10$^{-5}$ 
that the observed number of photons in the source cell is produced by a pure 
background fluctuation (about 4 $\sigma$ 
detection, Cruddace et al. 1988). The measured counts were then corrected for 
PSPC vignetting and the source parameters were 
determined by the maximum likelihood method ({\sc MAXLIK}). 
The sources which partially fell under the PSPC window support 
structure and which were missed by the detection routine 
have also been analyzed, and the vignetting-corrected count rates have been obtained 
in the different energy bands. 
Even though the above described routines have been extensively
applied in the past, 
we have further checked the reliability of all the detections 
by running the slide-cell detection algorithm in {\sc XIMAGE} 
(Giommi et al. 1991) in 
a similar way to that described in Fiore et al. (2000a) for all the sources 
of the present sample, obtaining similar results within the errors. 
Most of the sources are also present in the WGA catalogue 
(a point source catalogue generated from all {\it ROSAT} PSPC observations, 
see White et al. 1994). 
The agreement between the present results and those from the WGA 
makes us further confident about the findings which will be 
described in the following sections. 

Images were constructed in 4 energy bands: PI channels 11--41 (corresponding to 
0.1--0.4 keV, band {\sf a}), 52--201 (0.5--2 keV, {\sf b}), 
52--90 (0.5--0.9 keV, {\sf c}) and 91-201 (0.9--2 keV, {\sf d}), 
which correspond to S, H, H1 and H2 in Hasinger et al. (1998). 
The hardness ratios have been defined as follows: 
{\sf HR1=(b$-$a)/(b$+$a)} and  {\sf HR2=(d$-$c)/(d$+$c)}. 

The soft X--ray fluxes have been computed in the 0.5--2 keV energy range 
directly from the best-fit spectrum (when the statistics were high enough to 
allow a spectral modeling of the {\it ROSAT} counts) or by converting the 
count rates into fluxes under the assumption of a power-law slope with 
photon index 2.3 and Galactic absorption only. 

\begin{landscape}
\begin{table*}
\centering
\caption[]{HELLAS sources detected by \ROSAT}
\begin{tabular}{|l|cc|c|c|c|c|c|c|c|c|c|}
\hline
Source ID & RA & DEC & Class. & Count Rate & F$_{0.5-2~keV}$ & HR1 & HR2 & ROR \# & Off-axis & Exposure & $N_{{\rm H}_{\rm gal}}$ \\
 & (J2000) & (J2000) & & ($\times$10$^{-2}$ c/s) & ($\times$10$^{-13}$ cgs) & & & & (arcmin) & (ks) & \\
\hline
\#~20 & 01 40 14.7 & $-$67 48 55.2 & $\star$ & 0.20$\pm{0.04}$ & 0.22 & 0.159$\pm{0.200}$ & $-$0.252$\pm{0.194}$ & 300043p & 6.2 & 14.8 & \\
\#~45 & 03 15 47.5 & $-$55 29 5.2 & 1 & 1.24$\pm{0.06}$ & 1.6 & $-$0.170$\pm{0.031}$ & 0.158$\pm{0.046}$ & 701036p & 15.6 & 45.8 & 2.88 \\
\#~46 & 03 17 32.7 & $-$55 20 24.9 & 1 & 1.22$\pm{0.06}$ & 1.2 & 0.720$\pm{0.026}$ & 0.367$\pm{0.038}$ & 701036p & 20.7 & 45.8 & 2.88 \\
\#~53 & 04 37 11.9 & $-$47 31 43.2 & 1 & 2.36$\pm{0.24}$ & 3.4 & $-$0.513$\pm{0.032}$ & 0.216$\pm{0.074}$ & 701184p & 19.7 & 6.1 & 1.74 \\
\#~54 & 04 38 46.2 & $-$47 27 56.9 & 1 & 1.09$\pm{0.17}$ & 0.88 & $-$0.390$\pm{0.063}$ & 0.373$\pm{0.113}$ & 701184p & 22.2 & 6.1 & 1.78 \\
\#~57 & 05 20 49.0 & $-$45 41 31.2 & & 5.24$\pm{0.37}$ & 6.0 & 0.018$\pm{0.051}$ & 0.119$\pm{0.070}$ & 700057p & 12.6 & 4.4 & 4.09 \\
\#~65 & 06 46 38.0 & $-$44 15 34.8 & 1 & 4.92$\pm{0.32}$ & 5.9 & 0.654$\pm{0.050}$ & 0.008$\pm{0.066}$ & 300226p & 15.9 & 5.5 & 6.19 \\
\#~66 & 23 19 31.2 & $-$42 42 11.1 &1.8 & 2.16$\pm{0.22}$ & 3.0 & 0.164$\pm{0.053}$ & 0.107$\pm{0.069}$ & 700333p & 22.8 & 7.2 & 1.96 \\
\#~72 & 03 33 12.4 & $-$36 19 46.7 & BL Lac & 6.07$\pm{0.30}$ & 7.6 & $-$0.083$\pm{0.034}$ & 0.066$\pm{0.049}$ & 700921p-1 & 12.3 & 7.7 & 1.47 \\
\#~73 & 03 36 55.9 & $-$36 15 55.9 & RLQ & 2.71$\pm{0.30}$ & 1.7 & 0.021$\pm{0.091}$ & 0.110$\pm{0.114}$ & 700921p-1 & 41.0 & 7.7 & 1.40 \\
\#~75 & 03 34 6.9 & $-$36 03 55.3 & 1 & 0.19$\pm{0.05}$ & 0.20 & 0.874$\pm{0.112}$ & $-$0.211$\pm{0.281}$ & 700921p-1 & 7.7 & 7.7 & 1.37 \\
\#~84 & 13 36 39.2 & $-$33 57 52.4 & RadGal & 4.84$\pm{0.17}$ & 7.0 & 0.845$\pm{0.024}$ & 0.045$\pm{0.035}$ & 600268p-1 & 0 & 18.4 & 4.10 \\
\#~85 & 22 02 59.9 & $-$32 04 37.4 & 1 & 0.20$\pm{0.05}$ & 0.18 & 0.439$\pm{0.157}$ & 0.371$\pm{0.186}$ & 800419p-1 & 13.3 & 13.5 & 1.62 \\
\#~92 & 13 48 45.1 & $-$30 29 40.3 & 1 & 5.57$\pm{0.10}$ & 6.6 & 0.246$\pm{0.039}$ & 0.088$\pm{0.050}$ & 700907p & 13.3 & 8.2 & 4.39 \\
\#~103\_1 & 00 45 44.4 & $-$25 15 29.9 & 1.9 & 1.10$\pm{0.13}$ & 2.6 & $-$0.369$\pm{0.038}$ & 0.198$\pm{0.071}$ & 600087p-0 & 25.3 & 11.6 & 1.42 \\
\#~107 & 00 48 8.4 & $-$25 04 56.0 & & 0.32$\pm{0.06}$ & 0.38 & 0.583$\pm{0.079}$ & 0.280$\pm{0.106}$ & 600087p-0 & 14.5 & 11.6 & 1.52 \\
\#~124 & 00 27 9.5 & $-$19 26 16.0 & 1 & 6.18$\pm{1.72}$ & 3.1 & $-$0.320$\pm{0.250}$ & $-$0.120$\pm{0.450}$ & & $\dagger\dagger$ & 0.3 & 1.85 \\
\#~137 & 09 46 37.4 & $-$14 07 46.0 & 1 & 0.44$\pm{0.06}$ & 0.44 & 0.369$\pm{0.126}$ & $-$0.082$\pm{0.136}$ & 701458p & 17.8 & 18.6 & 4.10 \\
\#~147 & 20 42 52.9 & $-$10 38 33.0 & 1 & 3.69$\pm{0.41}$ & 3.7 & 0.385$\pm{0.068}$ & 0.431$\pm{0.081}$ & 701362p & 20.7 & 3.3 & 4.10 \\
\#~149 & 20 44 34.7 & $-$10 27 54.4 & 1 & 0.44$\pm{0.13}$ & 0.54 & 0.394$\pm{0.160}$ & $-$0.182$\pm{0.210}$ & 701362p & 16.5 & 3.3 & 4.20 \\
\#~150 & 13 05 32.9 & $-$10 33 15.9 & RLQ & 19.8$\pm{0.08}$ & 23.0 & 0.220$\pm{0.031}$ & 0.122$\pm{0.040}$ & 701195p & 0 & 3.2 & 3.33 \\
\#~151 & 13 04 33.0 & $-$10 24 37.6 & 1 & 0.61$\pm{0.17}$ & 0.62 & 0.152$\pm{0.172}$ & 0.158$\pm{0.227}$ & 701195p & 19.6 & 3.2 & 3.35 \\
\#~157 & 12 56 12.8 & $-$05 56 28.2 & 1 & 0.94$\pm{0.15}$ & 1.1 & 0.279$\pm{0.144}$ & 0.132$\pm{0.157}$ & 700305p-0 & 9.1 & 4.9 & 2.25 \\
\#~167 & 12 40 27.3 & $-$05 13 57.5 & 1 & 1.41$\pm{0.36}$ & 1.3 & 0.440$\pm{0.180}$ & $-$0.111$\pm{0.234}$ & 701012p & 13.8 & 1.3 & 2.28 \\
\#~172 & 02 42 1.2 & 00 00 26.6 & 1 & 0.64$\pm{0.12}$ & 0.88 & 0.133$\pm{0.143}$ & 0.261$\pm{0.172}$ & 150021p-2 & 9.9 & 5.5 & 3.45 \\
\#~174 & 02 42 11.0 & 00 02 2.2 & & 0.37$\pm{0.09}$ & 0.26 & 0.760$\pm{0.130}$ & 0.273$\pm{0.205}$ & 150021p-2 & 7.9 & 5.5 & 3.46 \\
\#~176 & 13 42 56.5 & 00 00 57.0 & & 0.95$\pm{0.07}$ & 1.2 & 0.074$\pm{0.041}$ & 0.135$\pm{0.056}$ & 701000p-1 & 17.3 & 27.8 & 1.91 \\
\#~185 & 05 15 15.7 & 01 09 19.5 & & 1.24$\pm{0.20}$ & 1.3 & 0.951$\pm{0.048}$ & 0.415$\pm{0.142}$ & 300352p & 7.7 & 3.1 & 10.9 \\
\#~190 & 16 52 38.4 & 02 22 3.9 & 1 & 0.44$\pm{0.13}$ & 0.34 & 0.765$\pm{0.156}$ & 0.385$\pm{0.256}$ & 701035p-1 & 5.7 & 3.2 & 5.75 \\
\#~201 & 16 49 59.2 & 04 53 37.6 & Cl. & 0.75$\pm{0.13}$ & 0.69 & 0.765$\pm{0.070}$ & 0.385$\pm{0.105}$ & 701611p & 18.2 & 8.1 & 6.35 \\
\#~209 & 23 27 29.2 & 08 49 28.0 & 1 & 0.52$\pm{0.13}$ & 0.41 & 0.579$\pm{0.166}$ & $-$0.313$\pm{0.217}$ & 600234p & 7.7 & 3.9 & 5.21 \\
\#~212 & 23 02 33.0 & 08 57 1.2 & ELG & 0.23$\pm{0.04}$ & 0.18 & 0.649$\pm{0.097}$ & 0.817$\pm{0.081}$ & 700423p & 11.3 & 18.6 & 4.91 \\
\#~229 & 23 31 54.7 & 19 38 29.0 & 1 & 0.64$\pm{0.14}$ & 1.5 & $\dagger$ & $\dagger$ & 201697h & 15.6 & 6.3 & 4.29 \\
\#~230\_1 & 15 28 48.1 & 19 38 52.7 & 1 & 0.94$\pm{0.36}$ & 0.60 & 0.453$\pm{0.261}$ & 0.262$\pm{0.333}$ & 180175p & 5.8 & 0.8 & 4.60 \\
\hline
\end{tabular}
\end{table*}
\end{landscape}

\begin{landscape}
\begin{table*}
\setcounter{table}{0}
\centering
\begin{tabular}{|l|cc|c|c|c|c|c|c|c|c|c|}
\multicolumn{12}{l}{{\bf {Table~1}} (continued).}\\
\hline
Source ID & RA & DEC & Class. & Count Rate & F$_{0.5-2~keV}$ & HR1 & HR2 & ROR \# & Off-axis & Exposure & $N_{{\rm H}_{\rm gal}}$ \\
 & (J2000) & (J2000) & & ($\times$10$^{-2}$ c/s) & ($\times$10$^{-13}$ cgs) & & & & (arcmin) & (ks) & \\
\hline
\#~230\_2 & 15 28 45.1 & 19 44 34.0 & & 3.69$\pm{0.72}$ & 4.3 & 0.935$\pm{0.063}$ & 0.505$\pm{0.158}$ & 180175p & 6.6 & 0.8 & 4.62 \\
\#~237 & 22 26 31.6 & 21 11 33.3 & 1 & 2.95$\pm{0.23}$ & 3.3 & 0.429$\pm{0.062}$ & 0.033$\pm{0.077}$ & 700856p & 14.0 & 6.4 & 4.44 \\
\#~239 & 14 17 16.9 & 24 59 17.3 & 1 & 0.49$\pm{0.08}$ & 0.95 & 0.002$\pm{0.123}$ & 0.107$\pm{0.157}$ & 700536p & 13.3 & 10.9 & 1.70 \\
\#~241 & 14 18 30.4 & 25 10 52.3 & Cl. & 3.51$\pm{0.19}$ & 6.3 & 0.438$\pm{0.046}$ & 0.337$\pm{0.052}$ & 600200p & 7.3 & 6.6 & 1.70 \\
\#~243 & 08 37 37.3 & 25 47 50.8 & 1 & 4.24$\pm{0.27}$ & 5.1 & 0.908$\pm{0.023}$ & 0.271$\pm{0.055}$ & 600200p & 10.9 & 6.6 & 3.64 \\
\#~246 & 08 38 58.3 & 26 08 18.6 & ELG & 3.33$\pm{0.28}$ & 4.5 & 0.937$\pm{0.024}$ & 0.781$\pm{0.043}$ & 600200p & 24.6 & 6.6 & 3.61 \\
\#~250 & 23 55 54.0 & 28 35 53.6 & RLQ & 1.86$\pm{0.11}$ & 2.2 & 0.617$\pm{0.049}$ & 0.244$\pm{0.057}$ & 200002p & 11.5 & 18.0 & 4.98 \\
\#~252 & 12 04 3.8 & 28 07 8.7 & Cl. & 5.51$\pm{0.17}$ & 7.9 & 0.467$\pm{0.027}$ & 0.287$\pm{0.030}$ & 700232p & 15.8 & 26.0 & 1.68 \\
\#~254 & 22 42 47.2 & 29 34 18.5 & & 0.41$\pm{0.11}$ & 0.42 & $-$0.160$\pm{0.173}$ & $-$0.396$\pm{0.238}$ & 701597p & 9.7 & 4.3 & 6.35 \\
\#~256 & 22 41 23.5 & 29 42 45.0 & & 6.23$\pm{0.42}$ & 7.3 & 0.642$\pm{0.051}$ & 0.297$\pm{0.064}$ & 701597p & 16.3 & 4.3 & 6.45 \\
\#~264 & 12 18 54.8 & 29 59 44.1 & 1.9 & 0.16$\pm{0.03}$ & 0.23 & $-$0.158$\pm{0.156}$ & 0.363$\pm{0.220}$ & 700221p & 15.8 & 21.6 & 1.70 \\
\#~265 & 12 17 52.3 & 30 07 4.3 & BL Lac & 13.9$\pm{0.26}$ & 16.0 & $-$0.375$\pm{0.010}$ & $-$0.023$\pm{0.019}$ & 700221p & 0 & 21.6 & 1.69 \\
\#~278 & 10 34 57.0 & 39 39 43.1 & & 0.52$\pm{0.16}$ & 0.27 & $-$0.149$\pm{0.224}$ & $-$0.338$\pm{0.309}$ & 700551p & 3.4 & 4.6 & 0.95 \\
\#~279 & 16 54 42.0 & 40 01 18.6 & & 2.00$\pm{0.22}$ & 0.98 & 0.443$\pm{0.072}$ & 0.282$\pm{0.091}$ & 700130p & 18.3 & 7.6 & 1.81 \\
\#~282 & 11 18 47.7 & 40 26 46.9 & 1 & 0.56$\pm{0.10}$ & 0.84 & $-$0.105$\pm{0.131}$ & 0.107$\pm{0.170}$ & 700801p & 3.3 & 6.5 & 1.92 \\
\#~283 & 11 18 14.0 & 40 28 34.3 & 1~red & 0.61$\pm{0.10}$ & 0.72 & 0.158$\pm{0.143}$ & $-$0.065$\pm{0.164}$ & 700801p & 4.3 & 6.5 & 1.91 \\
\#~288 & 12 19 23.2 & 47 09 42.6 & & 1.29$\pm{0.07}$ & 1.5 & 0.801$\pm{0.052}$ & 0.228$\pm{0.057}$ & 600546p & 9.9 & 25.7 & 1.15 \\
\#~290 & 12 19 52.1 & 47 21 0.0 & 1 & 0.34$\pm{0.04}$ & 0.72 & $-$0.142$\pm{0.082}$ & 0.177$\pm{0.113}$ & 600546p & 9.5 & 25.7 & 1.16 \\
\#~292 & 12 17 43.3 & 47 29 15.8 & & 2.99$\pm{0.12}$ & 4.0 & 0.524$\pm{0.040}$ & 0.173$\pm{0.037}$ & 600546p & 16.5 & 25.7 & 1.21 \\
\#~296 & 18 15 13.5 & 49 44 15.0 & & 0.10$\pm{0.03}$ & 0.09 & 0.825$\pm{0.118}$ & 0.081$\pm{0.218}$ & 300067p-1 & 11.6 & 18.0 & 4.34 \\
\#~300 & 10 32 15.6 & 50 51 12.0 & 1 & 0.22$\pm{0.05}$ & 0.14 & 0.353$\pm{0.126}$ & 0.551$\pm{0.136}$ & 701544p & 9.2 & 10.8 & 1.18 \\
\#~307 & 16 26 59.7 & 55 28 16.2 & Cl. & 6.49$\pm{0.57}$ & 3.9 & 0.476$\pm{0.076}$ & 0.330$\pm{0.081}$ & 701372p & 9.6 & 2.3 & 1.85 \\
\#~319 & 10 54 20.2 & 57 25 43.1 & 1.8 & 1.17$\pm{0.05}$ & 1.4 & 0.849$\pm{0.018}$ & 0.398$\pm{0.033}$ & 900029p-2 & 18.6 & 65.6 & 0.57 \\
\#~321 & 10 32 38.3 & 57 31 3.4 & & 0.17$\pm{0.02}$ & 0.54 & 0.659$\pm{0.110}$ & 0.116$\pm{0.147}$ & 900029p-2 & 9.9 & 65.6 & 0.56 \\
\#~364 & 14 38 22.0 & 64 31 17.9 & & 0.21$\pm{0.06}$ & 0.26 & 0.516$\pm{0.172}$ & 0.409$\pm{0.212}$ & 200069p & 14.5 & 7.4 & 1.68 \\
\#~375\_2 & 17 43 0.0 & 68 00 46.3 & & 2.75$\pm{0.33}$ & 1.1 & 0.432$\pm{0.074}$ & 0.229$\pm{0.099}$ & 999995p & 41.5 & 6.4 & 4.38 \\
\#~385 & 07 21 36.9 & 71 13 25.5 & 1 & 0.60$\pm{0.06}$ & 0.69 & 0.267$\pm{0.079}$ & 0.169$\pm{0.093}$ & 700210p & 7.1 & 20.7 & 3.84 \\
\#~387 & 11 01 48.8 & 72 25 44.1 & RLQ & 2.74$\pm{0.19}$ & 3.4 & 0.628$\pm{0.035}$ & 0.392$\pm{0.046}$ & 700872p & 24.3 & 13.1 & 3.16 \\
\#~389 & 11 06 16.6 & 72 44 10.5 & 1 & 0.65$\pm{0.08}$ & 0.73 & 0.038$\pm{0.092}$ & $-$0.012$\pm{0.116}$ & 700872p & 10.3 & 13.1 & 3.17 \\
\#~390 & 11 02 37.2 & 72 46 38.1 & 1 & 15.5$\pm{0.04}$ & 19.0 & 0.346$\pm{0.020}$ & 0.179$\pm{0.025}$ & 700872p & 22.4 & 13.1 & 3.36 \\
\#~392 & 07 41 44.6 & 74 14 41.6 & Cl. & 17.0$\pm{0.05}$ & 27.0 & 0.669$\pm{0.019}$ & 0.241$\pm{0.016}$ & 800230p & 0 & 8.8 & 3.49 \\
\#~393 & 07 42 2.4 & 74 26 21.9 & & 1.55$\pm{0.14}$ & 1.8 & 0.373$\pm{0.076}$ & 0.172$\pm{0.092}$ & 800230p & 11.4 & 8.8 & 3.52 \\
\#~394 & 07 43 12.0 & 74 29 34.0 & 1 & 9.53$\pm{0.36}$ & 11.0 & 0.235$\pm{0.029}$ & 0.056$\pm{0.037}$ & 800230p & 15.6 & 8.8 & 3.53 \\
\#~400 & 12 22 6.9 & 75 26 17.4 & Cl. & 1.01$\pm{0.12}$ & 1.1 & 0.671$\pm{0.101}$ & 0.404$\pm{0.116}$ & 700434p-1 & 7.8 & 7.4 & 2.97 \\
\hline
\end{tabular}
\pn 
{\small $\dagger$ HRI data; $\dagger\dagger$ RASS; Galactic $N_{\rm H}$ from Dickey \& Lockman (1990) in units of 10$^{20}$ cm$^{-2}$; 
Class: 1: broad-line AGN; 1~red: broad-line QSO 
with a red continuum; 1.8-2: narrow-line AGN; ELG: emission-line galaxy; 
BL Lac: BL Lac object; RLQ: radio-loud quasar; 
RadGal: radio-galaxy; Cl.: cluster; $\star$: star.}
\end{table*}
\end{landscape}

\subsection{ROSAT data analysis}
 
Each \ROSAT field has been analyzed searching for the soft X--ray counterpart 
of the \SAX HELLAS sources, assuming a cross-correlation radius of 100 arcsec. 
Six \ROSAT sources have been found at larger radii: five have been spectroscopically identified 
and the soft X--ray position coincides with the optical one within $\sim$ 30 arcsec. 
This evidence, coupled with the errors which can be associated to 
\SAX MECS pointing position reconstruction (mainly due to the 
unavailability of one star-tracker for a part of the observation, see paper~II for a full description 
of these problems), makes us confident that these associations are real. 
The sixth source belongs to a high-Galactic field where 
\SAX position reconstruction was not possible due to the absence of a bright known source 
in the same field of view as the HELLAS source. 

For the majority of fields only one \ROSAT  source is present in 
each \SAX error box. 
For the three cases in which more than one \ROSAT source is present, 
the one closest to the center of {\it BeppoSAX}  error box has been chosen. 
For these sources the soft X--ray emission is also associated to 
radio emission. 

Given the large range of both \ROSAT exposure times and off-axis angles distributions, 
the sensitivity limit is different from field to field. As a consequence, 
the number of spurious {\it ROSAT} - {\it BeppoSAX} associations has been derived by computing for each source 
the number of objects expected at its flux according to the 
0.5--2 keV integral source counts (Hasinger et al. 1998) and adopting a 
searching radius of 100 arcsec. 
The final number of spurious associations is therefore the summed contribution of 
the chance coincidences expected for each field. 
With this approach, we expect 2 associations by chance. 

{\sf 68} out of the 104 HELLAS sources have a {\it ROSAT} soft X--ray counterpart, while 
{\sf 36} sources went undetected. The 5--10 keV flux distribution for the 
sources detected by \ROSAT is not significantly different from that of the entire HELLAS sample 
(20 \% confidence level according to the Kolmogorov-Smirnov (KS) test).

\subsection{HELLAS sources detected by ROSAT}

Among the 68 sources detected by \ROSAT 
51 have been identified with extragalactic 
objects (see Table~1 for a comprehensive view on their soft X--ray properties), 
i.e. 35 Type~1 objects (including 4 radio-loud AGNs and 1 red quasar), 4 Type 1.8--2, 
2 emission-line galaxies, 
2 BL Lac objects, 1 radio-galaxy and 6 clusters. 
Moreover, one source is associated with a bright K star. 
All the radio-loud AGNs and the clusters, 
and about one third of the Type~1 objects have been identified through existing 
catalogues, while the remaining have been spectroscopically identified 
by our group (see La Franca et al. 2001). 
In several cases the \ROSAT--\SAX association has been used in the 
spectroscopic identification process. 
For what concerns the \ROSAT-optical association,  
we have considered the X--ray source physically associated with the 
optically identified object if the distance between the X--ray and the 
optical position is below 40 arcsec (which should take into account the 
dependence of the PSPC point-spread function with the off-axis angle and 
the possible error in aspect reconstruction). 
The {\it ROSAT} \ sources of the present sample have typical distances of 
about 15--20 arcsec from the optical position (which is consistent with the values found 
by Voges et al. 1999 for point-like sources). 
Only for two sources the optical and the {\it ROSAT} positions seem to be different: 
in one case the most likely explanation is that the X--ray emission comes from 
a group of galaxies (the likely counterpart of the HELLAS source) 
and not from a single object (which is confirmed by 
the presence of some optical emission-line objects in the BeppoSAX error box). 
In the other case, the X--ray source is clearly extended and identified with a Type~1.9 AGN.

\subsection{HELLAS sources undetected by ROSAT}

The off-axis angles and the exposure times distributions 
of the 68 detected sources are not statistically different -- according to the 
KS test -- from those of the 36 undetected sources. 

X--ray absorption is likely to play a major role in ``hiding'' a fraction of objects, thus 
making them extremely faint or even absent in the soft X--rays. 
This is confirmed by computing the average \SAX softness ratio (SR=(S$-$H)/(S$+$H), 
where S=1.3--4.5 keV and H=4.5--10 keV band) for the 
subsamples which are detected and undetected by \ROSAT, the former having 
SR = 0.037 (dispersion=0.363), the latter SR = $-$0.314 (dispersion = 0.424), 
this value corresponding to $N_{\rm H}$ $>$ 10$^{23}$ cm$^{-2}$ at any redshift, 
see Fig.~7 in Comastri et al. 2001 (paper~III). 
This is a further clear indication that absorption plays a major role 
in the soft X--ray detection/undetection of the HELLAS sources. 

Among the 12 sources which are spectroscopically identified 
without being detected by \ROSAT, 
we find 2 high-z Type~1 objects (z=0.953 and z=2.386), 3 emission-line 
galaxies, 1 red quasar and 6 Type 1.8-2. These objects, on the 
basis of the X--ray analysis performed with {\it BeppoSAX}\ , are likely to be affected by 
absorption, their mean softness ratio being $<$SR$>$ $\simeq$ $-$0.228 
($N_{\rm H}$ $>$ 10$^{23}$ cm$^{-2}$ at z $>$ 0.015).

\section{ROSAT results}

\subsection{Absorption}

The {\it BeppoSAX} 5--10 keV and {\it ROSAT} 0.5--2 keV fluxes are reported in Fig.~1. 
The dashed lines indicate the fluxes 
expected for a spectral slope of $\Gamma$=2.3 and $\Gamma$=1.6 (with the average Galactic absorption 
of the present distribution: $\sim$ 3$\times$10$^{20}$ cm$^{-2}$), 
and $\Gamma$=1.8 plus $N_{\rm H}$=10$^{22}$ cm$^{-2}$, from top to bottom. 
It appears clear that for a significant 
fraction of the objects intrinsic absorption and/or flatter 
spectral slopes are required in order to reproduce the 
observed flux ratio. It must be kept in mind that the values 
derived from Fig.~1 are only indicative and may be considered as lower limits 
on the presence of X--ray absorption, since any additional component in the 
{\it ROSAT} energy range would increase the soft X--ray flux, and that 
source variability between the observations with the two satellites 
may affect the results. 
However, since the amplitude of nuclear variability scales as (about) the inverse 
of the luminosity (Nandra et al. 1997), for the majority of the objects of the 
present subsample (with an average 0.5--2 keV luminosity of $\sim$ 10$^{44}$ erg s$^{-1}$) 
variability is unlikely to significantly affect the present results. 
\begin{figure}
\resizebox{\hsize}{!}{\includegraphics{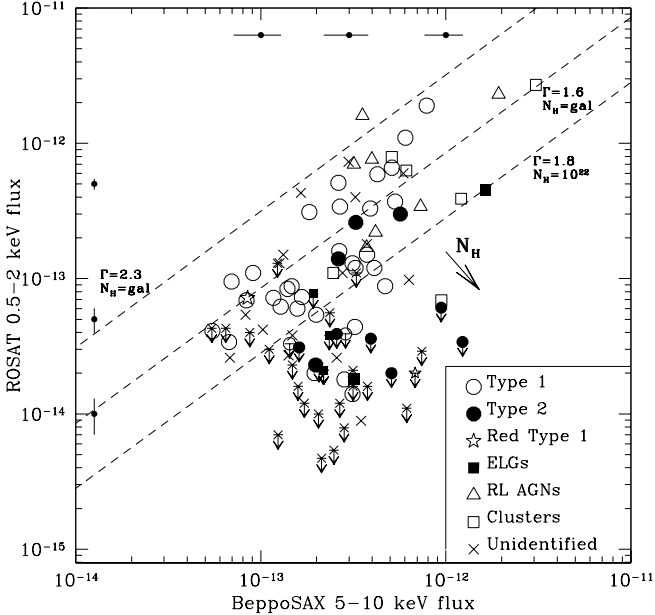}}
\caption{{\it BeppoSAX} 5--10 keV flux against {\it ROSAT} 0.5--2 keV 
flux for the HELLAS sources.  
The dashed lines indicate the expectations for $\Gamma$=2.3, 1.6 and 
Galactic absorption, and $\Gamma$=1.8 plus $N_{\rm H}$=10$^{22}$ cm$^{-2}$, from top to bottom. 
The errors associated to \SAX and \ROSAT fluxes are plotted separately (for 
three flux intervals).}
\label{fig1} 
\end{figure}

In order to provide a further evidence of the presence of strong absorption in the 
HELLAS sources, the upper limits for the sources which went undetected in the \ROSAT band 
are also plotted in Fig.~1. 
12 of these upper limits are relative to 
spectroscopically identified sources (cfr. $\S$ 3.4). 
It can be seen that the upper limits 
are extremely concentrated in the region 
characterized by high absorbing column densities: this is a further indication of the 
role of absorption in hiding the HELLAS sources, despite of their optical classification. 

\begin{figure*}
\centering
\resizebox{0.48\hsize}{!}{\includegraphics{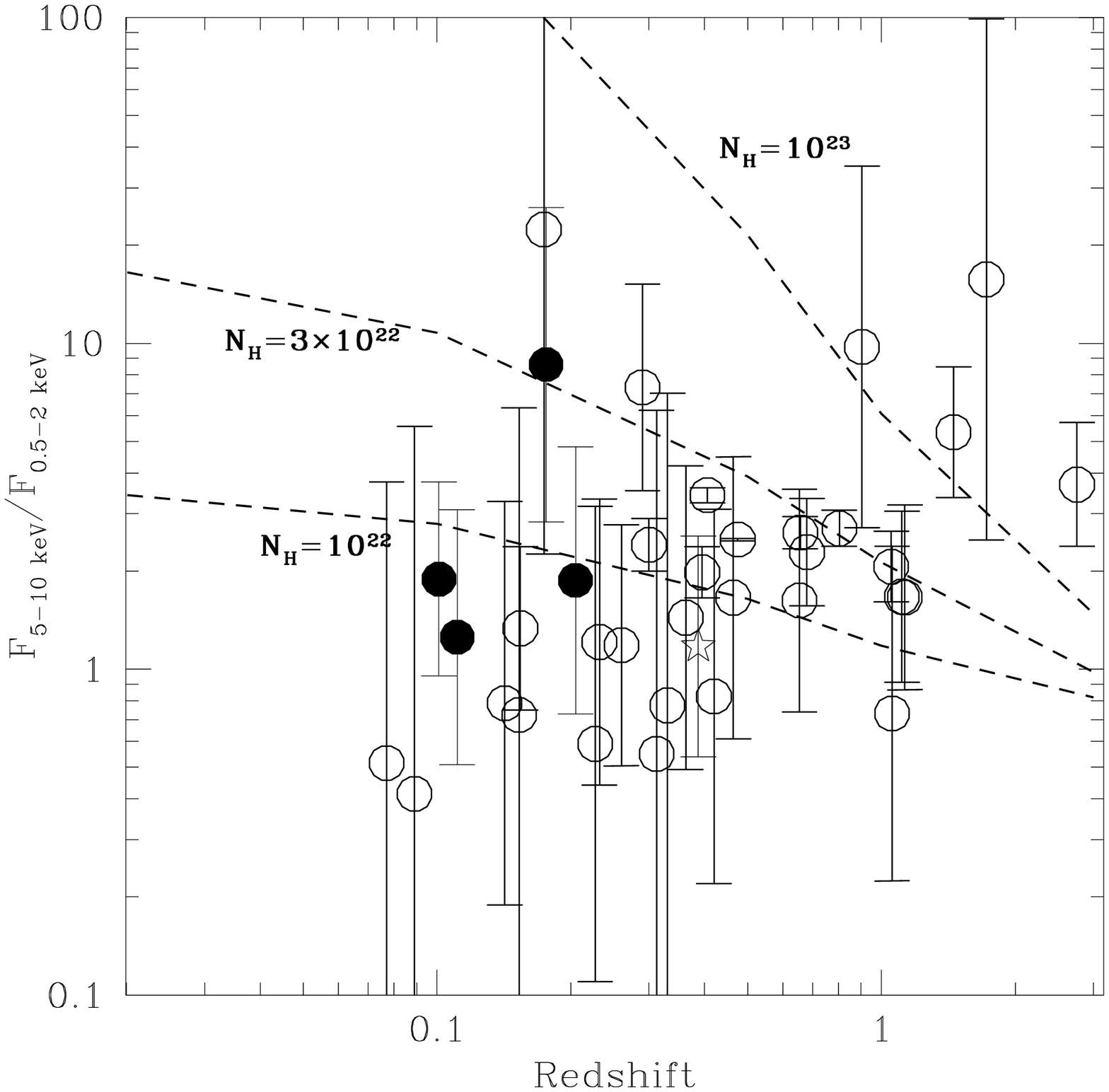}}
\resizebox{0.48\hsize}{!}{\includegraphics{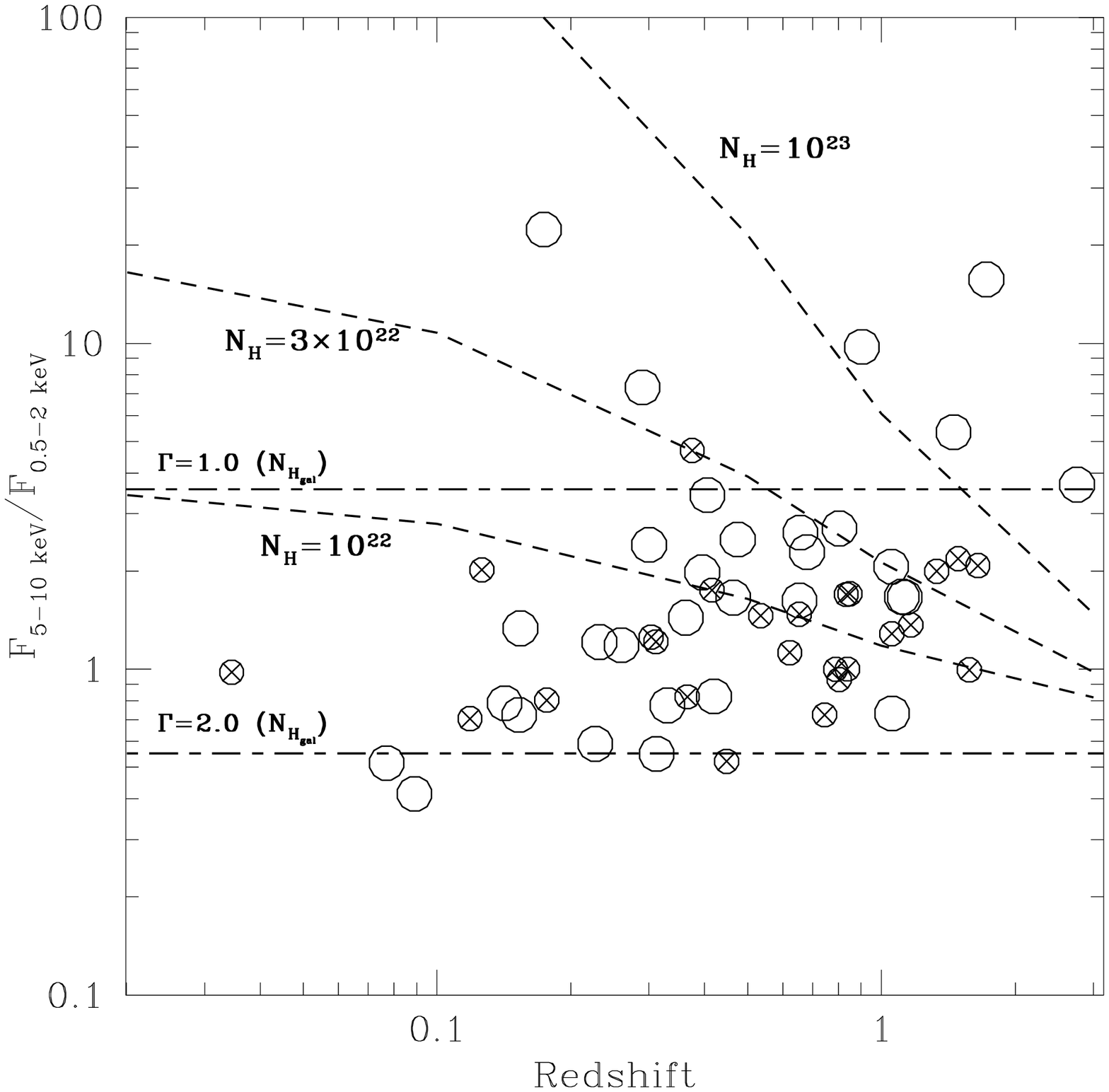}}
\caption{Left panel: {\it BeppoSAX} 5--10 keV to {\it ROSAT} 0.5--2 keV 
flux ratios as a function of redshift. Only the HELLAS sources 
classified as Type~1, Type~2 and the red quasar are plotted (symbols are as in 
the previous figure), along with three curves representing different column densities 
($N_{\rm H}$ = 10$^{22}$, 3$\times$10$^{22}$ and 10$^{23}$ cm$^{-2}$, 
assuming a power-law spectrum with $\Gamma$ = 1.8). 
Right panel: 5--10 keV to 0.5--2 keV flux ratios for the HELLAS (as in the previous panel, big open symbols) 
and the {\it ASCA} LSS broad-line objects (small crossed circles). 
The short-long dashed lines indicate the flux ratios expected for a $\Gamma$=1.0 and 2.0 power-law 
and Galactic absorption.}
\label{fig2ab} 
\end{figure*}

Similar results can be obtained by plotting the 5--10 keV to 0.5--2 keV flux 
ratios as a function of redshift (Fig.~2, left panel) for 
Type~1 and Type~2 sources. In order to provide an estimate of the 
absorption at the source redshift, three curves corresponding to Log$N_{\rm H}$ = 22, 
22.5 and 23 cm$^{-2}$ are also plotted. 
The underlying assumed spectrum is a power law with $\Gamma$ = 1.8. 
It is evident that about half of the sources require absorption in excess to 
10$^{22}$ cm$^{-2}$ and, among them, many are associated to broad-line objects. 
Viceversa, 3 out of 4 Type~2 AGNs are consistent with $N_{\rm H}$ $<$ 10$^{22}$ cm$^{-2}$. 
We also note a trend of increasing hardness with redshift, which is more pronounced for Type~1 
objects. 

A similar trend 
can be found by adding \ASCA  Large Sky Survey (LSS) broad-line objects 
(small crossed circles in Fig.~2, right panel) to the HELLAS ones. 
The increasing hardness with redshift, parameterized by the relation $F_{\rm hard}$/$F_{\rm soft}$ $\propto$ $z^{0.7}$, 
is particular evident at z $>$ 0.3 and could be ascribed to a 
flattening of the primary X--ray spectral slope at moderate/high redshift, as tentatively suggested by Vignali 
et al. (1999) for a small sample of high-z radio-quiet quasars. Even though the large majority of 
the objects of Fig.~2 (right panel) can be parameterized by a variety of spectral slopes ranging from $\Gamma$ = 1.0 to 
$\Gamma$ = 2.0 (the short-long dashed lines), 
the flux ratios of some sources are not well reproduced, 
clearly suggesting that absorption is a better explanation, 
in line with \SAX softness ratio predictions. 
A combination of these effects, absorption and flattening with redshift, is also plausible.

\subsection{Soft X--ray components}

The broad-band X--ray spectrum has been investigated by 
selecting a subsample of 12 HELLAS sources 
spectroscopically identified as broad-line objects, 
whose softness ratio (as a function of redshift) indicates 
the presence of substantial intrinsic absorption ($N_{\rm H}$ $>$ 5 $\times$ 10$^{22}$ cm$^{-2}$ 
at the source rest frame). 
For these sources we have calculated the 0.5--2 keV flux 
which is expected by extrapolating the \SAX flux in the \ROSAT band 
under the assumption of a $\Gamma$ = 1.8 power-law continuum 
absorbed by the $N_{\rm H}$ obtained by the softness ratio analysis 
with {\it BeppoSAX}\ .
These values have been subsequently compared to the observed 0.5--2 keV fluxes 
obtained directly from our analysis. 

Among the 12 Type~1 objects, 
5 sources have \ROSAT fluxes which agree with the 
predicted ones within 10 \%, while for 6 
of the remaining objects, in order to reproduce 
the observed \ROSAT flux, an additional component is strongly required, 
with a moderate/high fraction (from 25 to 65 \%) 
of the nuclear (5--10 keV) radiation being reprocessed in soft X--ray 
radiation and re-emitted. 
The remaining source is characterized by a soft X--ray flux higher than 5--10 keV flux, 
clearly pointing towards a prominent soft excess in this object. 

Similar values (from 5 to 35 \%) have been obtained for 
the 2 Type~2 objects which appear absorbed in \SAX and which 
are detected by \ROSAT. 

The true nature of the soft component is far from being clear: 
it could be due to nuclear photons 
spilling from a partial covering screen, or reflected by a warm/hot 
medium or, alternatively, it could be thermal emission from starburst 
regions and winds. At the present it is difficult to shed light on the nature of this 
component on the basis of these X--ray data only, 
unless {\it Chandra} or {\it XMM-Newton} follow-up observations 
will be performed.

\section{Comparison with the ROSAT Deep Survey sample}

In order to further investigate the soft X--ray properties and 
the nature of the hard X--ray-selected HELLAS sources, we compared the 
present sample with that of similar size extracted from the {\it ROSAT}  Deep Survey in the 
Lockman Hole (hereafter RDS, Hasinger et al. 1998; 
Schmidt et al. 1998; Lehmann et al. 2000), which is soft X--ray-selected 
and fully identified at a limiting X--ray flux of 5.5 $\times$ 10$^{-15}$ 
erg cm$^{-2}$ s$^{-1}$ (0.5--2 keV). The redshift distributions of the two samples are 
shown in Fig.~3. 
\begin{figure*}
\centering
\resizebox{0.48\hsize}{!}{\includegraphics{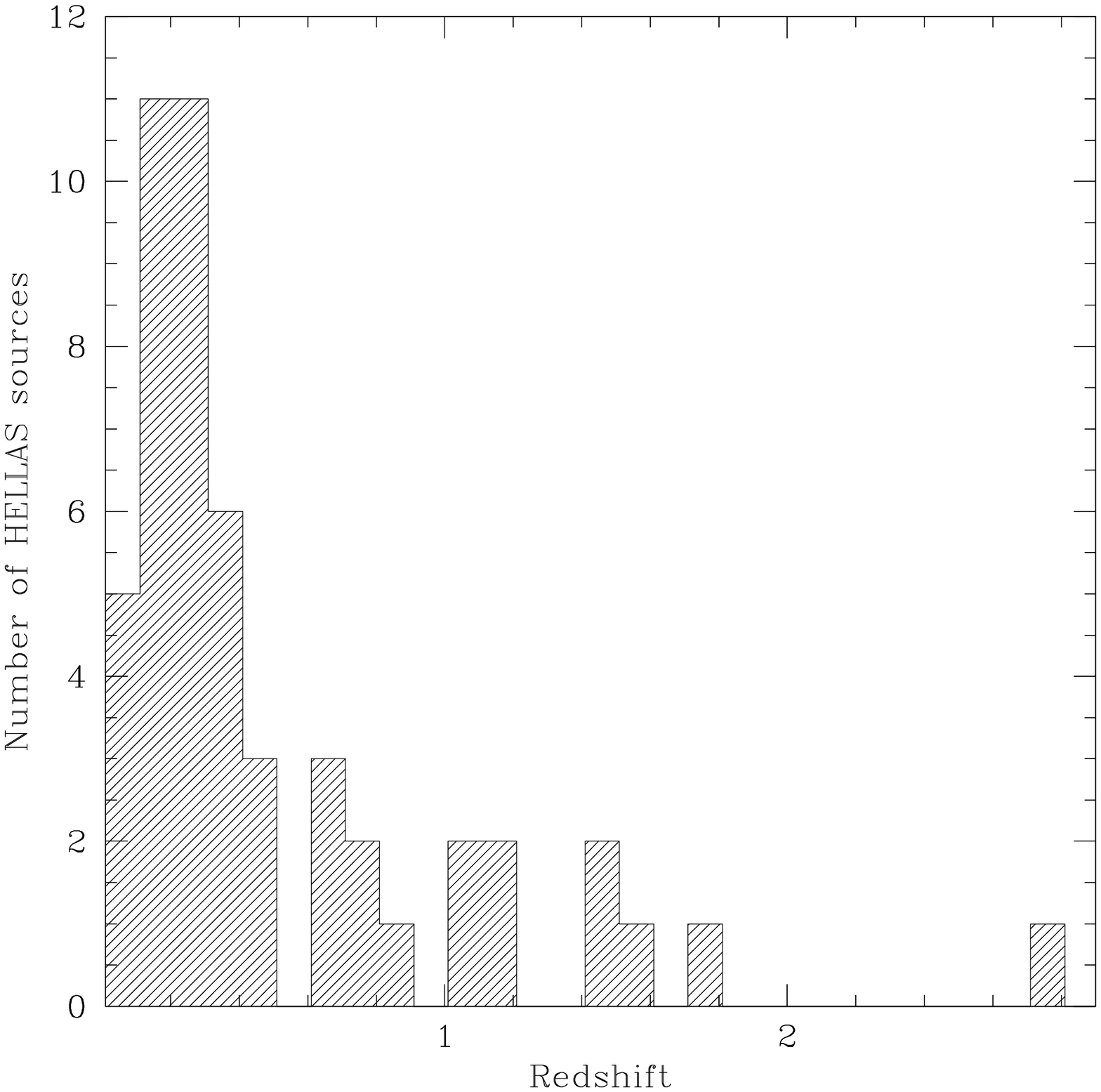}}
\resizebox{0.48\hsize}{!}{\includegraphics{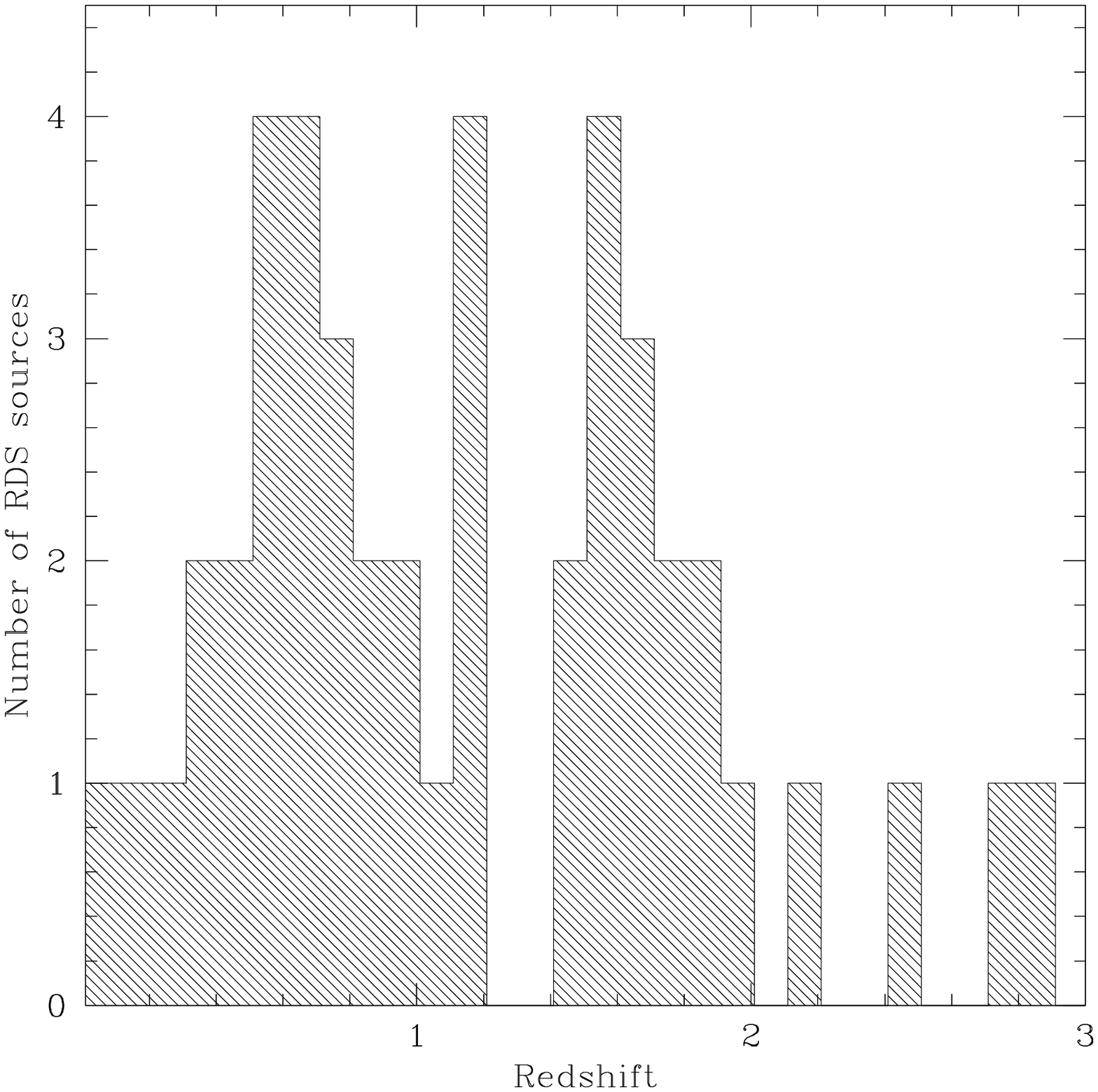}}
\caption{Redshift distribution for the HELLAS (left panel) and the RDS sources (right panel).}
\label{fig3ab} 
\end{figure*}

The properties of the two samples have been compared by means 
of hardness ratios HR1 and HR2 (cfr. Sect.~3.1). 
Figure~4 shows the plot of 
HR1 vs. HR2 for the HELLAS and the RDS sources 
(left and right panel, respectively).
\begin{figure*}
\centering
\resizebox{0.48\hsize}{!}{\includegraphics{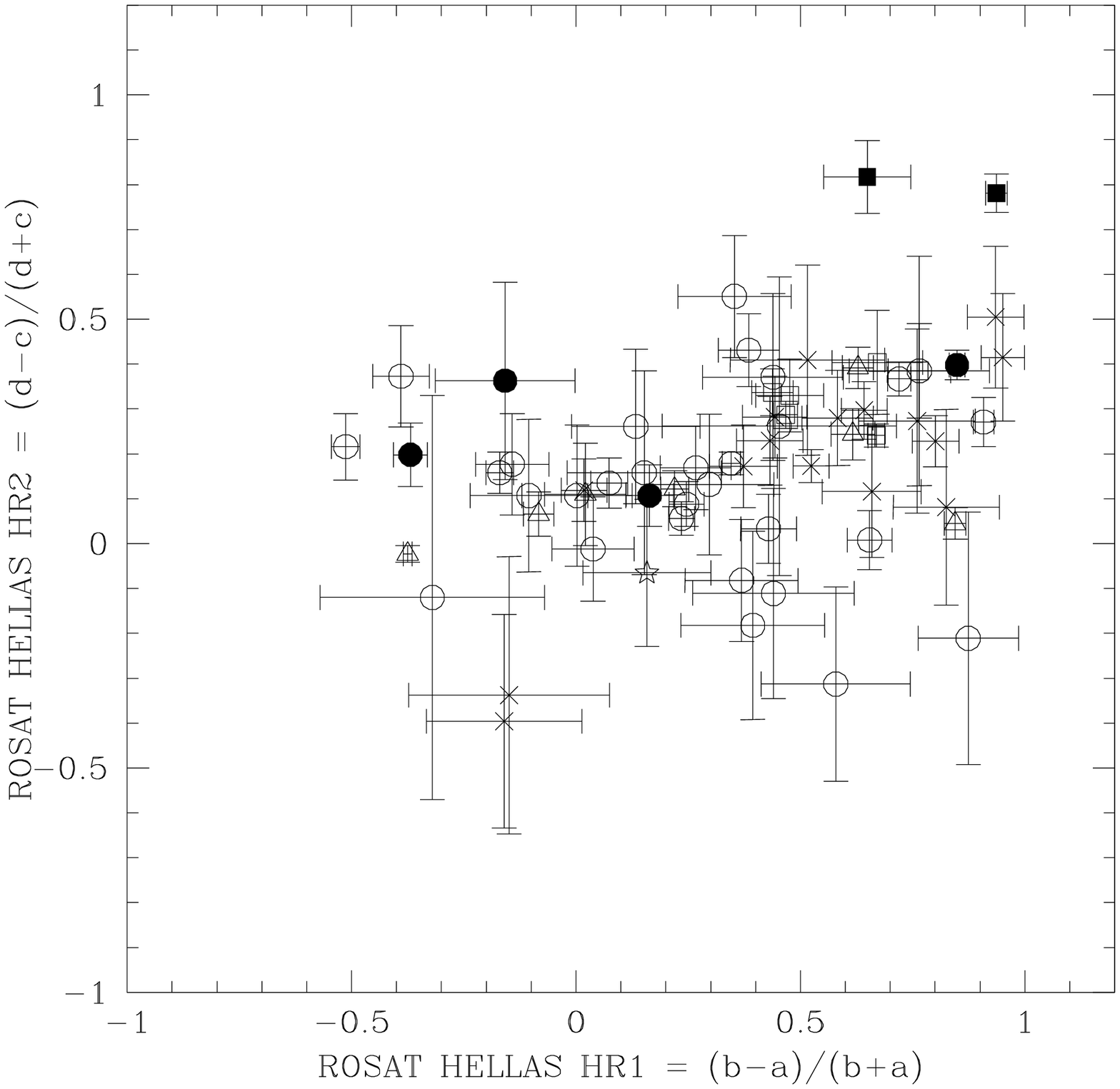}}
\resizebox{0.48\hsize}{!}{\includegraphics{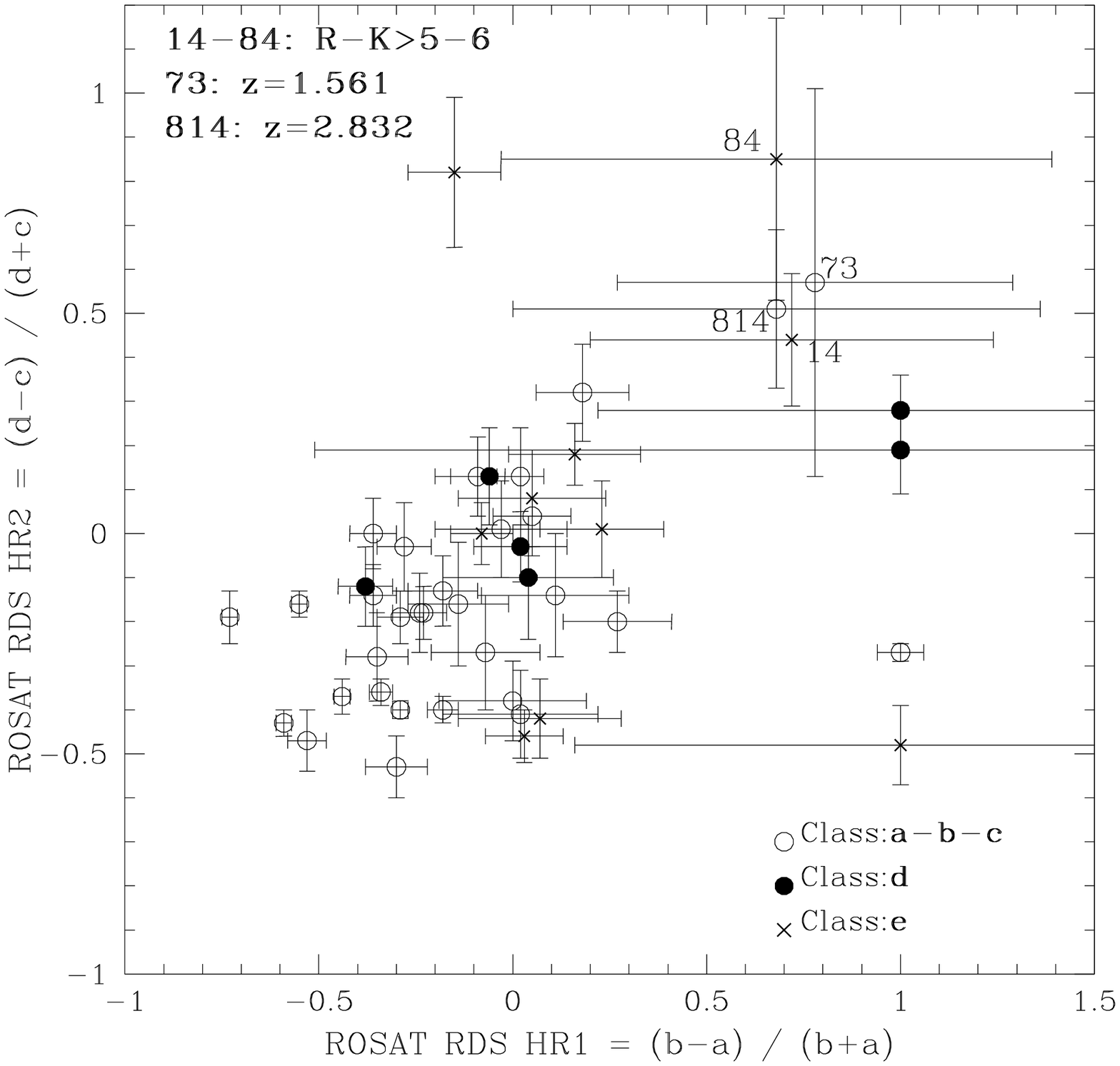}}
\caption{\ROSAT HR1 vs. HR2 for the HELLAS sources (left panel) and for the RDS sources (right panel).}
\label{fig4ab} 
\end{figure*}
It appears clear that most of the HELLAS sources populate 
the hardest region (HR1 $>$ 0, HR2 $>$ 0) of the diagram, while with a 
few exceptions the Lockman Hole sources are softer. 
It must be noted, however, that HR1 is very sensitive even to relatively small values 
of the Galactic absorption, that towards the Lockman Hole being of the order of 5.7 $\times$ 10$^{19}$ cm$^{-2}$, 
while the HELLAS sample spans a wide range of Galactic $N_{\rm H}$, from 5.7 $\times$ 10$^{19}$ cm$^{-2}$ 
(two sources detected in the Lockman Hole) 
to 1.1 $\times$ 10$^{21}$ cm$^{-2}$ (only one source having this $N_{\rm H}$), 
with an average column density of 2.6 $\times$ 10$^{20}$ cm$^{-2}$. 
This result in a shift of the HR1 distribution up to 0.5--0.6.

Therefore, since a meaningful comparison between HR1 for the two samples 
is not possible, a more reliable result 
can be obtained by comparing HR2, which is much less sensible to the Galactic 
absorption. Indeed, given the HELLAS $N_{\rm H}$ 
distribution, the effect of absorption is only marginal and 
could give rise to an increase in HR2 of only 0.05--0.1 
(which is within the statistical errors). 
The result is that the HR2 distribution of the HELLAS sample (Fig.~5, left panel) is intrinsecally 
different from that obtained for the Lockman Hole sources (Fig.~5, right panel), as 
confirmed by the Kolmogorov-Smirnov test (probability of 10$^{-6}$ that 
the two samples are drawn from the same parent population). 
The differences observed in the color-color diagram between the two samples 
are also evident in the color-redshift distribution. In particular, 
the HR2 of the HELLAS sources (Fig.~6, left panel) populates, on average, a 
harder region (upper right corner) 
of the diagram than the RDS sources (Fig.~6, right panel). 
\begin{figure*}
\centering
\resizebox{0.48\hsize}{!}{\includegraphics{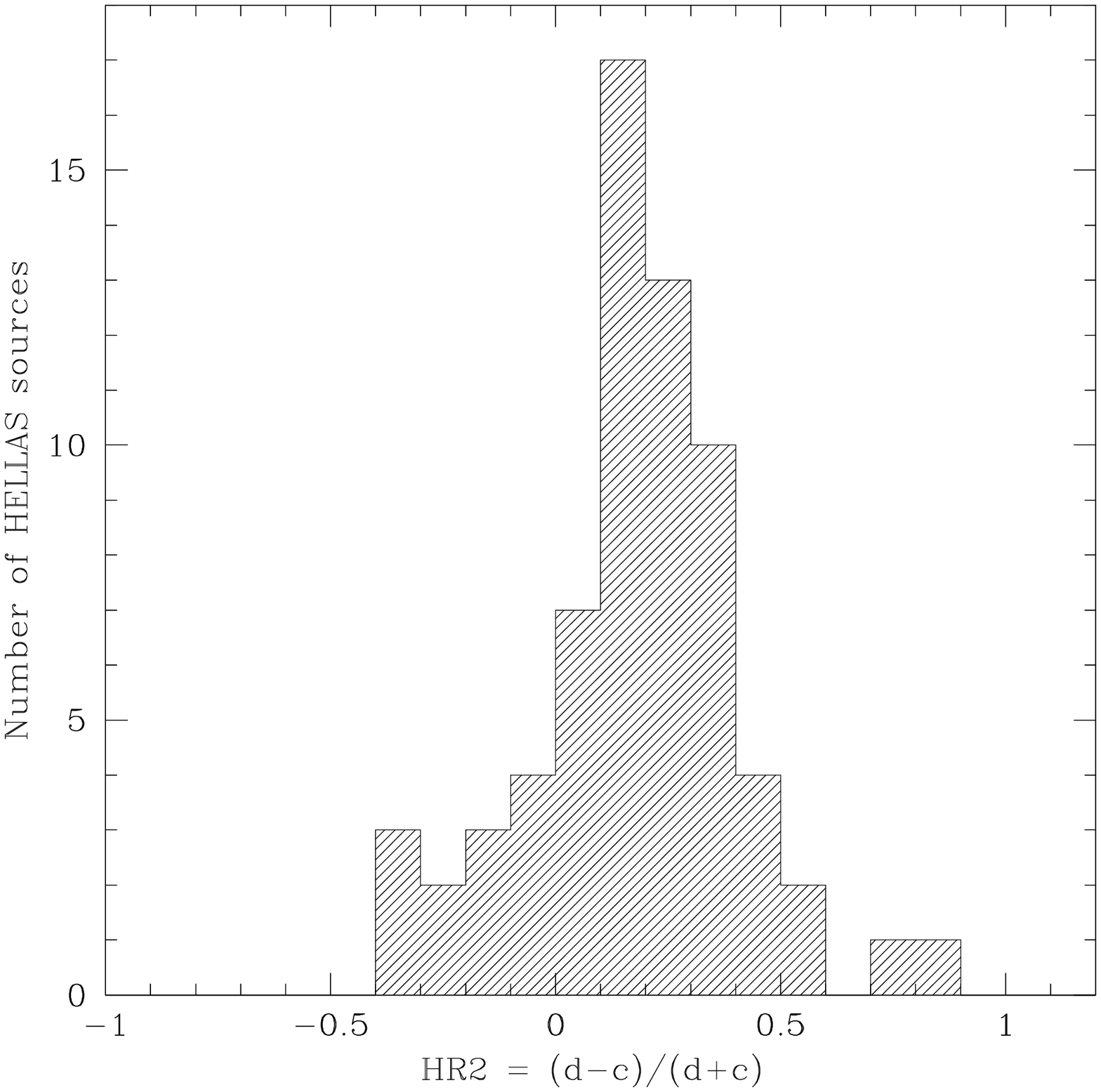}}
\resizebox{0.48\hsize}{!}{\includegraphics{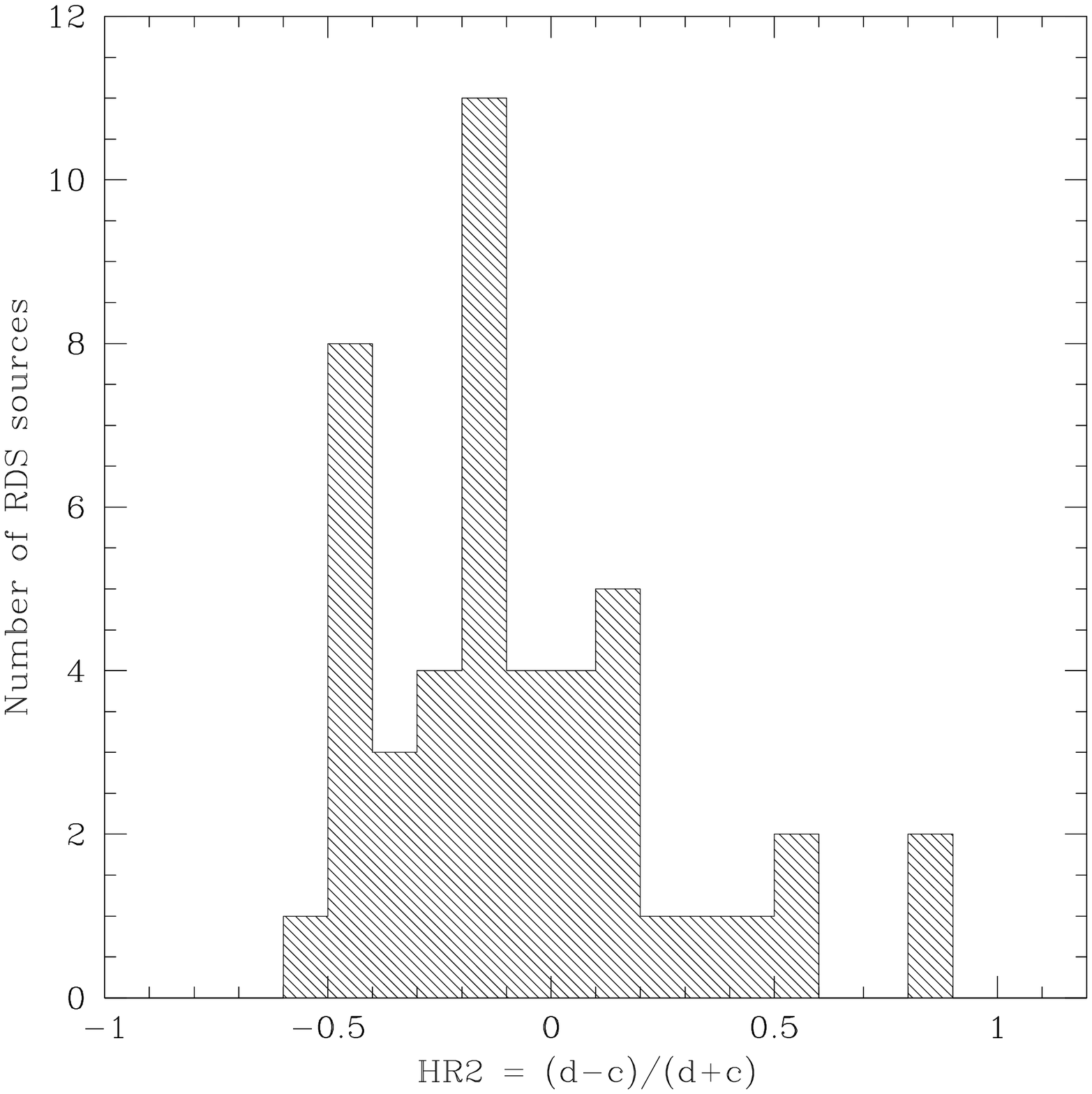}}
\caption{\ROSAT HR2 distribution for the HELLAS (left panel) and the RDS sources (right panel).}
\label{fig5ab} 
\end{figure*}
It is interesting to note that the variety of optical classification of hard sources 
is also found in the Lockman Hole sample (even though based on a smaller number of objects), 
where in the hardest part of the color-color diagram (Fig.~4, right panel) 
there are two Type~2 objects, two 
R$-$K$>$5--6 sources (candidate EROs, Extremely Red Galaxies; see Lehmann 
et al. 2000 for a detailed discussion) and two Type~1 objects, whose redshifts 
(1.561 and 2.832) and hard X--ray colors seem to confirm 
\SAX findings on X--ray absorbed Type~1s. 

All the previous results indicate that 
hard X--ray selection provides a significant fraction of absorbed objects, 
characterized by varied optical properties and classifications (broad-line AGNs, 
narrow-line AGNs, emission-line galaxies). 
X--ray absorbed objects are however present also in soft X--ray 
selected samples, even though at a lower level.  
Most interesting, a large fraction of hard X--ray-selected objects 
is present in \ROSAT: 
the obscured constituents of the XRB progressively show up going to 
fainter fluxes and harder energy ranges.  
\begin{figure*}
\centering
\resizebox{0.48\hsize}{!}{\includegraphics{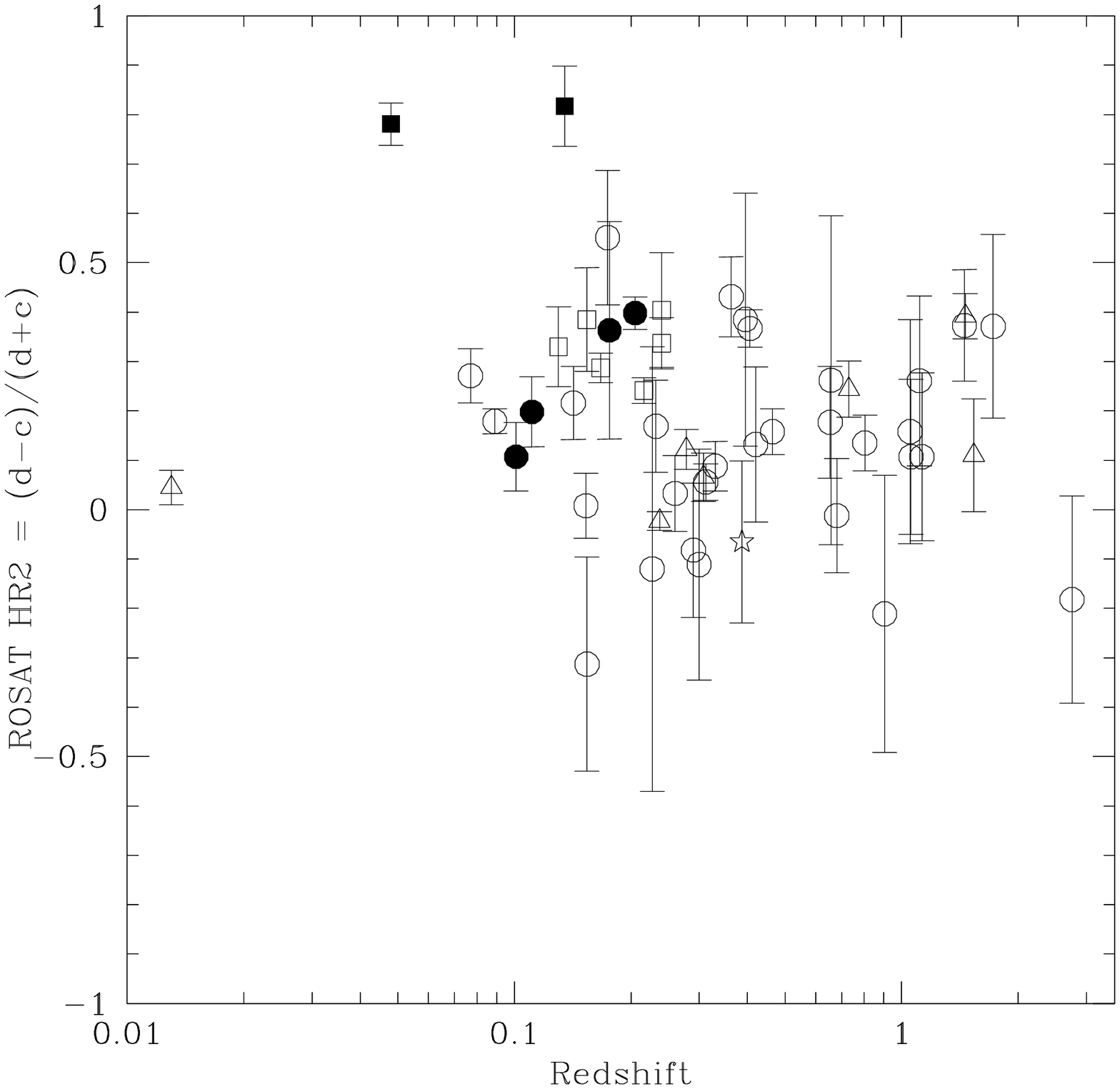}}
\resizebox{0.48\hsize}{!}{\includegraphics{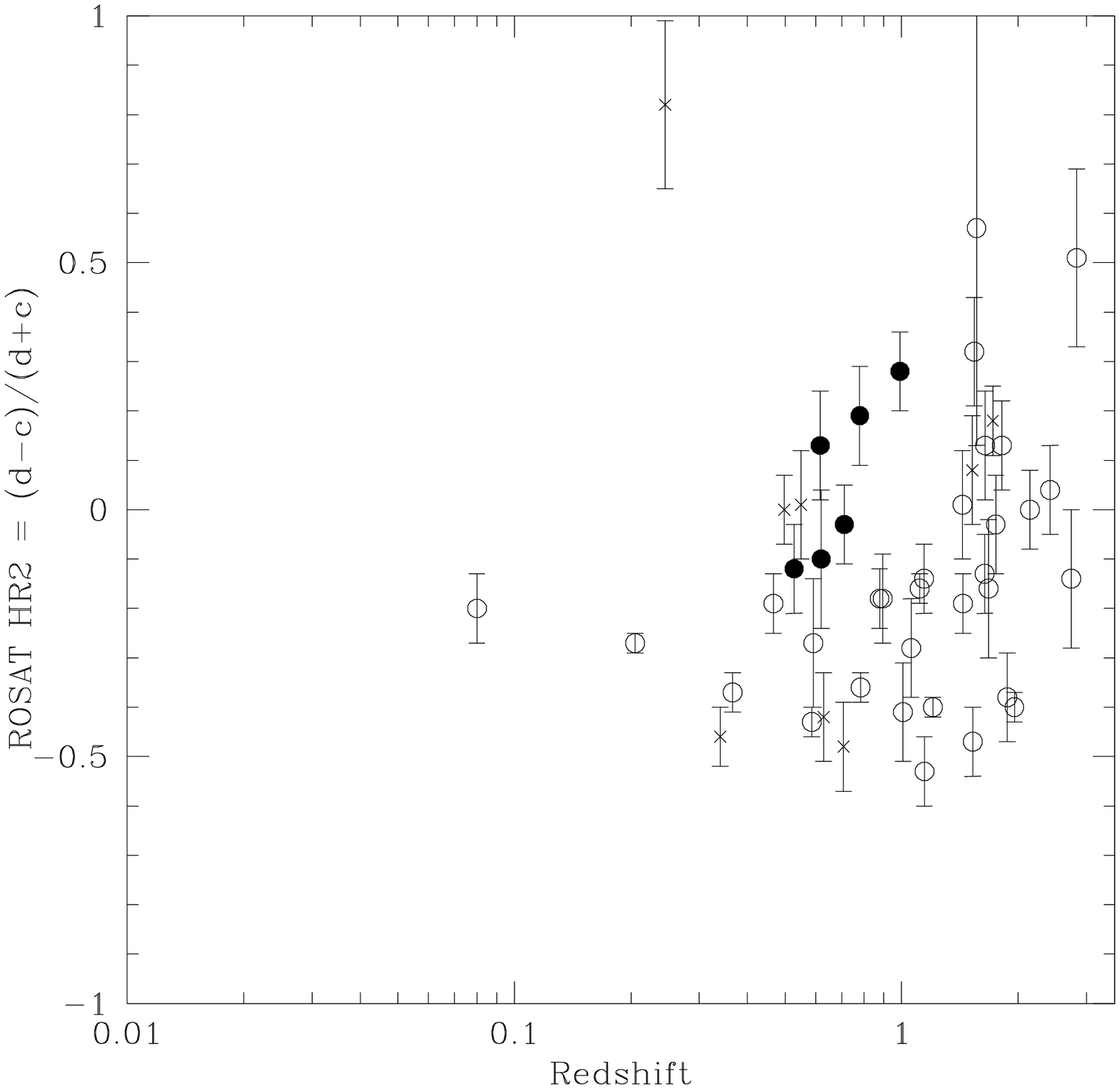}}
\caption{HR2 vs. redshift for the HELLAS sources (left panel) 
and the Lockman Hole objects (right panel).}
\label{fig6ab} 
\end{figure*}

\section{Discussion and conclusions}

The recent hard X--ray surveys performed by {\it ASCA}\ , \SAX and 
{\it Chandra} have revealed that the population responsible for the 
hard X--ray background has quite varied 
multiwavelength properties. According to AGN synthesis models (i.e. Comastri 
et al. 1995), these sources must be characterized by a 
spectral energy density spanning a wide range of luminosities and 
absorption column densities, in order to reproduce both the XRB spectrum 
and the source counts in different energy ranges. 
In particular, the energetically dominant contribution comes from sources 
around the knee of the X--ray luminosity function 
($L_{\rm X}$ $\sim$ a few 10$^{44}$ erg s$^{-1}$ at z=1) and with 
absorbing column densities of the order of 10$^{23}$ cm$^{-2}$ (Comastri 2000). 
These objects, the so-called ``QSO~2'', i.e. quasars with optical narrow-emission lines, 
have been extensively searched, but, at present, 
only a handful of candidates have been found (Ohta et al. 1996; 
Akiyama et al. 2000; Della Ceca et al. 2000). 
Results from recent hard X--ray surveys and the ones presented in this paper 
indicate that 
X--ray absorption is not correlated with the optical reddening, hence 
the classical statement of unobscured Type~1 -- obscured Type~2 objects 
is far from being always valid. 
Indeed, Type~1 AGNs characterized by hard X--ray colors, likely to be 
absorbed by column density of the order of 5 $\times$ 10$^{22}$ cm$^{-2}$ or greater, 
have been found by both \SAX  and \ASCA  surveys (Akiyama et al. 2000), 
and their role in contributing 
to the XRB may be exactly the same as the postulated Type~2 QSOs. 

The emerging picture is that the zoo of the hard X--ray-selected sources 
is characterized by more complex properties than previously thought 
(Comastri et al. 2000; Vignali et al. 2000). 

Another interesting result comes from the relatively high number of 
HELLAS hard X--ray selected sources revealed by \ROSAT. 
This is due to a combination of effects, the most important being 
the high sensitivity of \ROSAT  also at faint flux levels and the presence of soft X--ray emission 
also in strongly obscured AGNs. 
The most important implication is that the same highly absorbed 
sources responsible for a sizeable fraction of the hard XRB emit in 
the soft X--rays (Giommi et al. 2000; Giacconi et al. 2001), where they 
confirm their hardness. 
In some cases it was found that this emission is enhanced, with respect 
to the extrapolation at low energies of the absorbed higher energy component, 
by Compton downscattering or through a thermal component. 
For a subsample of HELLAS sources it has been possible to provide an 
estimate of the relative fraction of this scattered component with respect to the nuclear flux, 
spanning from a few to 65 per cent. 

Although this component is not energetically dominant (indeed 
it is not present in the XRB synthesis model by Comastri et al. 1995), 
it could be important when comparing surveys performed in different energy 
ranges and, most important, at limiting fluxes differing by one (or more) 
order of magnitude. This has interesting consequences also in the computation of 
the X--ray luminosity function for the sources responsible for the XRB, since 
the contribution from this additional component is taken into account as it was 
of nuclear origin, not due either to reprocessing of the primary radiation or to thermal 
emission. 

Remarkably, the \ROSAT analysis indicates that 
a significant number of HELLAS sources are characterized by hard X--ray colors 
also in soft X--rays. 
The broad-band analysis of the sources which are not detected by \ROSAT 
(upper limits in Fig.~1) suggests that truly hard sources 
(where the soft X--ray emission is extremely faint or totally absent) 
do exist. 
There are evidences that such sources have also been detected by {\it XMM-Newton} 
in the Lockman Hole (Hasinger et al. 2001). 

The comparison of the color-color properties of the present sample with 
those of the ROSAT Deep Survey (Hasinger et al. 1998) in the Lockman Hole 
confirms that we are sampling the hard tail of 
the distribution of the sources responsible for the XRB, 
where the hardness of the spectrum may be ascribed to a flat continuum or, more 
convincingly, to large amounts of X--ray absorption, or to a combination 
of both the factors. 
Even though the HELLAS sources are harder in soft X--rays than the RDS objects, 
it is straightforward to note a continuity of properties between the two samples, 
with a optically varied population of hard sources in both.

\begin{acknowledgements}

This research has made use of the NASA/IPAC Extragalactic
Database (NED) which is operated by the Jet Propulsion Laboratory,
California Institute of Technology, under contract with the National
Aeronautics and Space Administration. 
The authors wish to thank G. Zamorani, G. Matt and P. Giommi for useful 
suggestions and discussion, and the anonymous referee for prompt and constructive comments. 
This work is partly supported
by the Italian Space Agency, contract ARS--99--79 and by the Ministry
for University and Research (MURST) under grant COFIN--98--02--32. 

\end{acknowledgements}

\end{document}